\documentclass[12pt,preprint]{aastex}

\shorttitle{VLBI Observation of Cyg X-3 in State Transition}
\shortauthors{Kim et al.}

\begin{document}


\title{VLBI Observation of Microquasar Cyg X-3 during
an X-ray State Transition from Soft to Hard in the 2007 May-June Flare}

\author{\sc Jeong-Sook Kim$^{1,2}$,
            Soon-Wook Kim$^{2,7}$,
            Tomoharu Kurayama$^{3,4}$,
            Mareki Honma$^5$,
            Tetsuo Sasao$^6$,
            Sang Joon Kim$^1$}
\affil{$^1$ School of Space Science, Kyunghee University,
            Seocheon-dong, Giheung-si, Gyeonggi-do, 446-701, Republic of Korea;
            evony@kasi.re.kr.\\
       $^2$ Korea Astronomy and Space Science Institute,
            776 Daedeokdaero, Yuseong, Daejeon 305-348, Republic of Korea;
            skim@kasi.re.kr.\\
       $^3$ Graduate School of Science and Engineering, Kagoshima University,
            1-21-35 Korimoto, Kagoshima, Kagoshima 890-0065, Japan \\
       $^4$ Center for Fundamental Education, Teikyo University of Science,
            2525 Yatsusawa, Uenohara, Yamanashi 409-0193, Japan \\
       $^5$ National Astronomical Observatory of Japan, 2-21-1 Osawa,
            Mitaka, Tokyo 181-8588, Japan \\
       $^6$ Yaeyama Star Club, Ookawa, Ishigaki, Okinawa 904-0022, Japan
       $^7$ Corresponding author.}
\altaffiltext{*}{corresponding author}
\begin{abstract}

   We present a radio observation of microquasar Cyg X-3
during an X-ray state transition
from ultrasoft to hard state in the 2007 May$-$June flare using
the VLBI Exploration of Radio Astrometry (VERA) at 22 GHz.
   During the transition, a short-lived mini-flare of ${\lesssim}3$ hr
was detected prior to the major flare.
   In such a transition, a jet ejection is believed
to occur, but there have been no direct observations to support it.
   An analysis of Gaussian fits to the observed visibility amplitudes
shows a time variation of the source axis, or a structural change,
during the mini-flare.
   Our model fits, together with other multiwavelength observations
in the radio, soft, and hard X-rays,
and the shock-in-jet models for other flaring activities at GHz wavebands,
suggest a high possibility of synchrotron flares during the mini-flare,
indicative of a predominant contribution from jet activity.
   Therefore, the mini-flare with an associated structural change is
indicative of a jet ejection event
in the state transition from ultrasoft to hard state.

\end{abstract}

{\it Online-only material: color figure}

\section{Introduction}

   Cyg X-3 is an X-ray binary that was first discovered in X-rays \citep{Gia67}.
   A short orbital period of 4.8 hr \citep{Par72,Bec73} and
a massive Wolf$-$Rayet donor star \citep{van96,Koch02} make Cyg X-3
a tight high-mass X-ray binary system.
   The nature of the compact object is
still a matter of debate, but there has been growing evidence
of a black hole \citep[e.g.,][and references therein]{Shrader10}.
   Cyg X-3 is the brightest X-ray binary in the radio and
has displayed frequent flaring activity
since the discovery of a giant flare in
1972 \citep{Gre72,Waltman94,Waltman95,Waltman96}.
   \citet{Waltman94} identified three types of variability in
the 1988$-$1992 light curves at 2.25 and 8.3 GHz:
frequent minor flaring activity of $<$1 Jy,
intermediate flares ($>$1 Jy), and major flares ($>$10 Jy).
   In addition, there are pre-flare quenched states of tens of mJy
prior to the flares, and flickering activity during quiescence.
   In very long baseline interferometry (VLBI) observations,
clearly resolved jet structures on tens of milliarcsecond (mas) scales
have been observed  \citep{Newell98,Mio01,Mil04,Tudose07,Tudose10}.
   Even in the observations with unresolved structure, model
fits indicate expanding jet-like structures
\citep[e.g.,][]{Gel83,spen86,Mol88,Sch95,Sch98}.
    However, the VLBI observations have mostly been carried out during major flares,
in particular after the peak, and rarely performed in the rise to the peak
mainly due to the slow reaction time of the VLBI facilities
after the triggering of flares.
    Therefore, it is important to probe the rise phase to understand
the evolution of flares and associated initiation of jets.

   Cyg X-3 exhibits the canonical X-ray states such as low/hard state,
very high state, intermediate state, high/soft state, ultrasoft state,
and quiescence.
   However, the X-ray spectra in Cyg X-3 are complex and significantly
different from other X-ray
binaries \citep[e.g.,][]{MR00,Fender04,S08,Hja09,kol10}.
   In Cyg X-3, the X-ray hardness intensity diagram (HID),
a plot of X-ray flux versus X-ray color (hardness),
does not show a typical hysteresis unlike other black hole X-ray
binaries \citep[][and references therein]{kol10}.
   \citet{Fender04} predict that a relativistic ejection is expected
to occur in a transition to a harder X-ray state across the ``jet line",
while soft states do not seem to produce jets.
   Based on further analysis of the HID, \citet{kol10} conclude that
Cyg X-3 crosses the jet line and a major radio flare occurs
when there is a transition from a ``hypersoft state"
(or canonical state of the ultrasoft state) to a flaring, harder X-ray state.

   On 2007 June 1, Cyg X-3 was found in a radio flare at 1$-$30 GHz with
the RATAN-600 radio telescope \citep{Tru07}.
   The radio flare reached $\sim$1.6 Jy at 1 GHz and
3.9 Jy at 30 GHz the next day, and declined to $<$1 Jy at 1$-$30 GHz
within the next two to three days (Figure 1).
   A few days prior to the radio observation,
the hard X-rays (15$-$50 keV) rapidly increased with the hardening
in the soft X-rays (see the third and bottom panels in Figure 1).
   After a soft state observed since 2007 April \citep{Krimm07}, an extreme soft,
or ultrasoft state was observed with $INTEGRAL$ during
May 21$-$26 (JD 2,454,242$-$2,454,247; \citep{Soldi07,Bec07}.
   During May 27$-$29 (JD 2,454,248$-$2,454,250),
the onset of an X-ray state transition was observed with $INTEGRAL$
and six days later during
June 5$-$12, it reached a hard state \citep{Bec07}.
   Our VLBI observation was carried out
after $INTEGRAL$ reported Cyg X-3 in its ultrasoft state at 3$-$20 keV
\citep{Soldi07}.
   In this paper, we present a VLBI observation of
the 2007 May-June flare of Cyg X-3 in the rise phase
during a state transition from ultrasoft to hard state.

\section{VERA Observation, Data Analysis and Results}

   Cyg X-3 was observed at 22 GHz for $\sim$9 hr,
during ${\sim}$15$-$24 UT on 2007 May 29,
with VLBI Exploration of Radio Astrometry (VERA).
   VERA consists of four 20 m antennas located at
Mizusawa, Iriki, Ogasawara and Ishigaki-jima.
    The baseline length of VERA ranges
from 1019 km (Iriki$-$Ishigaki) to 2270 km (Mizusawa$-$Ishigaki),
and the expected angular resolution at 22 GHz is $\sim$1.2 mas,
depending on the projected baseline in the $uv$-plane.

   A left-handed circular polarization (LCP) was received
because the LCP signal is only available in VERA.
   The received signals were sampled with 2$-$bit quantization
and filtered to 16 MHz bandwidth using VERA digital filter bank 
(Iguchi et al. 2005).
   Since the recording rate of VERA DIR2000 recorder is 1024 Mbit s$^{-1}$,
we can record $16~{\rm MHz~bandwidth} \times 16~{\rm IFs}$ in Nyquist sampling
($16~{\rm MHz} \times 2~{\rm samples~Hz^{-1}} \times 2~{\rm bit~sample^{-1}} \times 16~{\rm IFs}
= 1024~{\rm Mbit~sec^{-1}}$).
   Of 16 IFs, 15 were assigned to Cyg X-3
and one IF was assigned to another target,
W75N \citep{Kim2013},
separated by $2.^{\circ}03$ from Cyg X-3.
   All VERA antennas have a facility of dual-beam receiving system
\citep{Kawaguchi00} for simultaneously observing a target and
reference source within $2.^{\circ}2$.
Since a short-duration flare of Cyg X-3 which we present in this paper was
readily detected in the fringe search (see below),
we did not carry out the phase-referencing VLBI technique.
   The data correlation processing was carried out on the
Mitaka FX Correlator \citep{Chi91} at the Mitaka campus of the
National Astronomical Observatory of Japan (NAOJ).

   The VLBI data analysis was performed using
the Astronomical Image Processing System
\citep[AIPS;][]{Greisen2003}.
   For amplitude calibration,
the flux density scale was set using system temperatures measured with
the chopper-wheel method \citep{Ulich76} and aperture
efficiency\footnote{http://veraserver.mtk.nao.ac.jp/restricted/CFP2010/status10.pdf}
depending on the elevation.
    In our VERA observation, we measured the power of a hot load
and a normal sky, corresponding to the ambient and sky temperature.
For the bandpass calibration,
we alternately observed BL Lac for 10 minutes and Cyg X-3
for 70 minutes in a cycle of 80 minutes,
which corresponds to the recording time of a tape of VERA's recorder.
   For the intrinsic variability of BL Lac, the intraday variability of
amplitude is a few percent \citep[e.g.,][]{Quirrenbach92}.
   Therefore, the calibration is free from the intrinsic variability of BL Lac.
   Fringe-fitting was carried out
separately for Cyg X-3 and the calibrator source, BL Lac.

   We detected fringes of Cyg X-3 with a threshold of the
signal-to-noise ratio of 4 in the AIPS task FRING.
   The visibility amplitude from the detected fringes for
each baseline is shown by the black crosses in Figure 2.
   All VERA stations had no systematic and technical problems
during our observation.
   In the Ishigaki and Ogasawara stations,
the system temperature highly varied up to
$\sim$1,500 and $\sim$3,000 K due to bad weather before 20:00 UT,
but became stabilized down to $\sim$500 and $\sim$200 K
after $\sim$20:00$-$21:00 UT, respectively.
   In the Iriki station, $\sim$200$-$250 K was maintained
except the rapid variation before 16:00 UT and $\sim$17:30$-$20:00 UT.
   The system temperature in the Mizusawa station remained fairly
constant, $\sim$150$-$200 K, during the observation.
   We excluded bad data from the data analysis for
system temperatures higher than $\sim$800 K or whenever the
system temperature rapidly changed by 50\% or more within tens of minutes.
   Therefore, Cyg X-3 was detected mainly at the Mizusawa$-$Iriki baseline
before 21:00 UT, while no detection was made at all baselines during
$\sim$17:30$-$20:00 UT.
   There is a short-duration flare (hereafter, “mini-flare”)
for 21:00$-$23:37 UT.
   The system temperature in all stations
became stabilized after 21:00 UT.
   Therefore, the observed mini-flare is independent of systematic effects
such as the antenna system temperature and source elevation
that began to decrease after 19:00 UT.
   Note that the mini-flare is detected on all baselines,
indicating that the mini-flare is
an intrinsic phenomenon of the target source.
   The peak flux density is greater than 1 Jy on
shorter baselines such as
the Iriki$-$Ogasawara and Iriki$-$Ishigaki baselines,
while it is less than 1 Jy on longer baselines
such as the Ogasawara$-$Ishigaki and Mizusawa$-$Ishigaki baselines.

    From the variation of the visibility amplitudes presented in Figure 2,
we infer that the detected fringes are not from a point source.
   The change of visibility amplitudes during the mini-flare
in Figure 2 is intrinsic to the source Cyg X-3,
and reminiscent of, for example,
few hour-scale small flares reported by \citet{Newell98}.
    For the time-variation in the visibility amplitude during the flares,
they suggested that
structure variation was seen.
    They indeed produced a sequence of images from the 20-minute snapshots,
implying a 0.3c jet event
that clearly confirmed the change in the source structure during the flares.
   To investigate whether our observations of Cyg X-3 
indicate temporal structural variation, we performed the following two tests.
   First, we tested an image by performing a hybrid mapping method
for all the detected fringes assuming no structural time variation, and
compared the corresponding model visibility amplitudes,
shown by the blue dashed lines
in Figure 2, with the observed visibility amplitudes.
   In the Mizusawa$-$Ishigaki baseline,
the model visibility amplitude over the observing time
(the blue dashed line in Figure 2)
is
larger than the observation
(the black crosses), indicating
that the model is too compact compared to the real size of Cyg X-3.
   Therefore, such a simple model without time variation
does not seem to well represent the observation.

   The second test we performed was an elliptical Gaussian fit
to the observed visibility amplitude.
   Due to the limitation in $uv$ coverage with
only six baselines,
we cannot directly produce a hybrid mapping image for a time-scale short
enough to trace a change of source structure along with the variation of
flux density.
   We checked for phase closure during observations
of Cyg X-3 and obtained a value near zero within ${\pm}10^{\circ}$,
which justifies adopting a single Gaussian fitting
to the visibilities.
   We fitted the visibility amplitude directly
in the $uv$-plane without making images.
   The model fitting was carried out using
Caltech's Difmap program.
   We adopted a single elliptical Gaussian component with a time bin of
10 minutes during the mini-flare, starting from 21:14 UT.
We obtained a flux density, major axis, minor axis, and
position angle of
the
major axis for each time bin from the model fit.
   The orange solid lines in Figure 2 are the visibilities
produced based on the model-fit results.
   In Figure 3, we present the
model-fit results showing
total flux density,
intrinsic source sizes of major and minor axes and position angle.
   The model flux density for each baseline,
shown by the orange solid lines for each time bin,
is in good agreement with the observation (Figure 2).
   Error analysis for the total flux density, the sizes of major and minor
axes and the position angle were estimated using Difwrap \citep{Lovell00}.
   The reduced ${\chi}^2$ for the model fits with time bins of
10 and 20 minutes are ${\lesssim}0.82$.
   The model fit with a time bin of 30 minutes
has a reduced ${\chi}^2 {\lesssim} 1.1$,
approximately equivalent to $1 {\sigma}$ \citep[e.g.][]{Dodson03}.
   The time-evolving features of axes and position
angles for the time bins of 10, 20 and 30 minutes
are consistent with each other.
   In the following discussion,
we exclude the first data point in Figure 3
because it is a marginal detection,
and the last data point
because the time bin is very short ($<8$ minutes).
   The major axis kept increasing since 22.5 UT,
${\sim}1$ hr before the peak flux density.
   In the two mini-flares observed by \citet{Newell98},
the extended structures (jet expansion) appeared
${\sim}1$ hr prior to the peak of each flare,
proportional to the increase in flux density
\citep[see Figures 4 and 5 in][]{Newell98}.
   Similarly,
the increase of the major axis size in Figure 3 may indicate
a structural change, or jet ejection, during the mini-flare
due to synchrotron emission.
   The position angle of the major axis is relatively stable
with time,
particularly after ${\sim}21.75$ UT (the bottom panel in Figure 3).
   This position angle of ${\sim}50{\pm}10^{\circ}$ is
in agreement with the observations of \citet{Newell98} and \citet{Sch95}.
   In the following section,
we discuss the possible causes of the structural change.

\section{Discussion and Conclusion}

   In this paper,
we present a short-lived ($\lesssim$3 hr) flaring event, or mini-flare
during an X-ray state transition from ultrasoft to hard state
in the early phase of the 2007 May$-$June flare.
   No radio observation was reported prior to the peak of a few Jy
except our VLBI observation.
   To make matters worse, no hard X-ray observation was made
during most of the 2007 May-June flare.
   However, it is clear that
a few days before our observation
the hard X-ray flux of $Swift$/BAT rapidly increased
after its $\sim$12.5 day low flux quenched state  (Figure 1).
   At the same time, the soft X-ray flux ($RXTE$/ASM)
decreased and a few days later
the radio flux was observed near the peak.
   It is well known that the strong correlation of
the radio and soft X-ray emission disappears immediately before
the flaring periods \citep{watanabe94,Chu02}.
   In flaring states, there is a correlation between
the radio and hard X-ray emission,
while the soft X-ray is anti-correlated with the two wavebands \citep{Mc99}.
   Therefore, at the time of our observation, the 2007 May$-$June flare was
probably already in the early rise.
   To confirm this,
we checked the multiwavelength light curves
of a few previous flares.
   In most of the major ($>$10 Jy) and intermediate ($<$10 Jy) flares,
the radio fluxes at GHz wavebands increase rapidly
from a few hundred mJy to a few Jy
when the $Swift$/BAT flux increases
from $<$0.005 to $\sim$0.02 counts s$^{-1}$ cm$^2$
\citep[$Swift$/BAT Web site; RATAN Web site\footnote{http://www.sao.ru/cats/$\sim$satr/XB/CygX-3/};
e.g.,][]{Tru08}.
   Therefore, at the time of our observation,
the 2007 May$-$June flare was most likely in its early rise phase
since the $Swift$/BAT flux was $\sim$0.02 counts s$^{-1}$ cm$^2$ (Figure 1).
   In Cyg X-3, small-scale flares on a time-scale of a few hours or less
have been observed during low-level activity of $<$600 mJy at 5$-$43 GHz
\citep[e.g.,][]{Newell98,Marti00,Ogley01,Mil09}, and
associated jet ejections were confirmed in some cases
\citep{Newell98,Marti00}.
   However, a flare on a similar time scale prior to
major or intermediate flares, such as the mini-flare of $\sim$1 Jy
that we have presented in this paper, has not been reported.

   Due to the absence of published X-ray spectra
during the radio flare of 2007 May-June
($\sim$JD 2,454,250$-$2,454,256; see Figure 1),
it is not clear whether Cyg X-3 underwent an intermediate
or very high state prior to the observed hard state
\citep{Bec07}.
   \citet{kol10} find that there are two types of flares:
flares with hard X-rays and without hard X-rays,
and the latter does not reveal
any jet-like structure \citep[also see][]{Fender04}.
   They point out that the first two $INTEGRAL$ observations
in 2007 May (JD 2,454,241 and JD 2,454,246) during a quenched state
correspond to the hypersoft or ultrasoft state (Figure 1).
   Therefore, a further transition to a harder state in
2007 May$-$June would result in crossing the jet line in HID.
   From Figure 1, it is clear that Cyg X-3 already changed
its state from the ultrasoft to a harder state
with rapidly increasing hard X-rays of $Swift$/BAT
during our observation.
   Therefore, the structural change in our observation (Figures 2 and 3)
indicates that the mini-flare is plausibly produced
by a jet ejection in the early phase of the 2007 May$-$June flare.
   Furthermore, \citet{Tru07} point out that,
 prior to the major flare in 2007 May$-$June,
the observations of $Swift$/BAT in mid- to late 2007 April
\citep{Krimm07}
and $INTEGRAL$ in 2007 May 21
\citep{Soldi07}
clearly indicate a  possible jet ejection event.

 What can be a cause of the mini-flare?
   The most appropriate model to account for such a mini-flare
would be the so-called shock-in-jet model \citep{MG85}.
   The model is an analytic, parameterized shock model describing
the synchrotron emission associated with relativistic jets
to account for the radio and near-infrared light curves
of flaring activities in quasars and microquasars
\citep{Turler00,Turler11}.
   The model can reproduce the variable jet emission in microquasars,
very similar to that of extra-galactic jets in quasars,
suggesting that the physical characteristics of the relativistic jets
are independent of the black hole masses
\citep{Turler07}.
   As for small-scale flares in other microquasars, for example,
\citet{Turler04} reproduced a series of flaring events of
a few hours observed at 2.2 ${\rm {\mu}m}$ and 2$-$6 cm
during a plateau state in the microquasar GRS 1915+105.
   In the case of Cyg X-3,
the model accounts for a series of small-scale, tens of minutes to an hour,
flaring activities in the decay phase of a half-day long radio flare
(${\lesssim}500$ mJy) at 1.4$-$43.3 GHz observed in 2002 January 25
\citep{Mil09}.
    The model can also reproduce the major and intermediate flares of Cyg X-3
by adopting a different time-scale, peak flux density 
and frequency of each flaring activity
at 2.2 ${\rm {\mu}m}$ and 2.25$-$15 GHz
\citep[e.g.,][]{Turler11},
suggesting that the same phenomena are at work.
    The difference is that the brighter flares peak at lower frequencies
and have longer time-scales than fainter flares, which is
consistent with the formation of shocks further downstream in the jet
\citep[for further discussion, see][]{Mil09,Turler11}.

   In our observation, the model fits with
a single elliptical Gaussian component
are in good agreement with the observed visibility amplitudes.
   The model fits indicate the variation of axis size
during the mini-flare, similar to
the previous observations of the two mini-flares
at different flux scales, both of which included jet imaging
\citep{Newell98}.
   A possibility of jet ejection is also supported by
other multiwavelength observations in the radio, soft and hard X-rays
during the 2007 May$-$June flare
\citep{Tru07,Krimm07,Soldi07,Bec07,kol10}.
   In particular, it has been suggested that the jet ejection is highly
expected during a state transition from soft to harder X-ray state
\citep[e.g.,][]{kol10, Fender04}.
   The shock-in-jet model
for microquasars suggests
that various flares with different time scales can be described by
a synchrotron flare
\citep{Turler00,Turler04,Turler11,Mil09}.
     Therefore, together with such observations and models
for flaring activities in Cyg X-3 and other microquasars,
our model fits suggest
a high possibility of synchrotron flares and
an
associated jet ejection
during the X-ray state transition from ultrasoft to hard state
in the 2007 May$-$June flare.
    The observation of jet activity in
this particular state
transition has been a long-awaited issue.
   We present
observational evidence of such an expected
jet ejection event in the state transition for the first time.
The next challenge is to image a variety of jet ejection events with the
combined array of VERA and Korean VLBI 
Network\footnote{http://kvn-web.kasi.re.kr/en/en{\_}index.php}
which will provide better UV coverage and
allow us to make snap shot images of
rapidly variable sources.

\acknowledgments
We thank the anonymous referee
for constructive suggestions and a thoughtful review of the manuscript
that has helped to substantially improve the contents of this paper.
   We thank Katsunori Shibata at NAOJ for arranging the dynamic scheduling.
   We thank Jim Lovell and Kotaro Niiuma for providing Difwrap and for useful comments on it.
   We also thank Makoto Inoue and Kiyoaki Wajima for comments on Difwrap.
   S.J.K. acknowledges partial supports from
the Korea Science and Engineering Foundation (R01-2008-000-20002-0),
a WCU Grant (No. R31-10016),
and a grant from the Kyung-Hee University in 2013.

\clearpage

\clearpage
\onecolumn
\clearpage
\begin{figure}
\epsscale{0.85}
\plotone{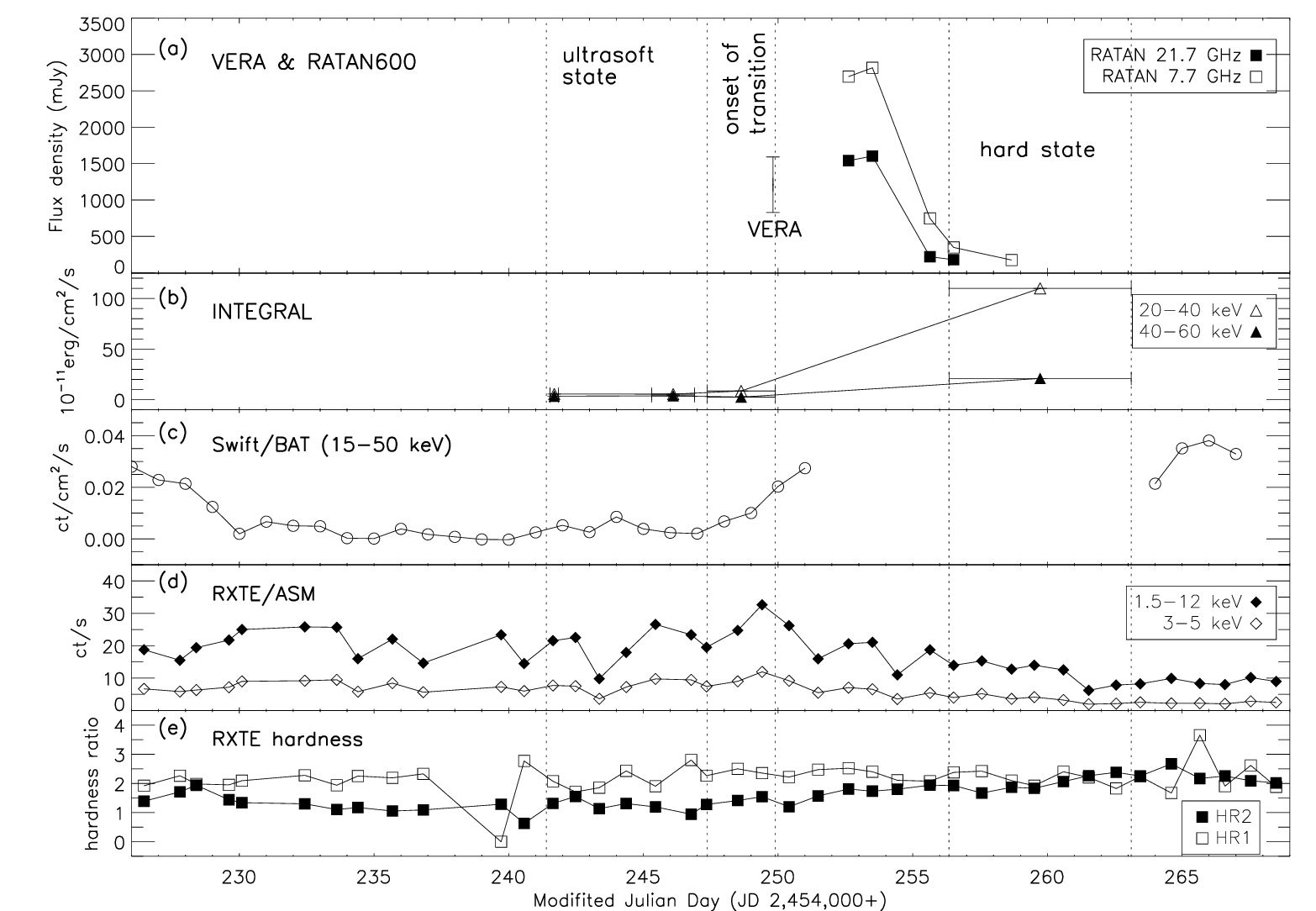}
\caption{Radio and X-ray light curves of the 2007 May$-$June  flare.
   Cyg X-3 light curves are presented for (a) radio flux density at 7.7 and 21.7 GHz
observed with RATAN-600 telescope on May 31, June 1, 3, 4 and 6 \citep{Tru07},
and the flux density
from VERA obtained with model fits, shown by
the vertical bar (see the text; Figures 2 and 4);
(b) hard X-rays on May 21, 25 and 27; June 5, 8, and 11
at 20$-$40 and 40$-$60 keV observed with the imager IBIS/ISGRI
of $INTEGRAL$ adopted from Tables 1 and 2 in \citet{Bec07};
(c) daily monitored hard X-rays at 15$-$50 keV observed with $Swift$/BAT adopted from
http://swift.gsfc.nasa.gov/docs/swift/results/transients/CygX-3/;
(d) daily monitored soft X-rays at 1.5$-$12 keV and 3$-$5 keV observed with $RXTE$/ASM; and
(e) the corresponding $RXTE$/ASM hardness ratio of
HR1 ($=$3$-$5 keV/1.5$-$3 keV) and HR2 ($=$5$-$12 keV/3$-$5 keV)
adopted from http://xte.mit.edu/ASM{\_}lc.html.
The observed three X-ray spectral states are presented in panel
(a) and the corresponding time intervals for each X-ray spectral state are
shown by the vertical dotted lines in all panels.
\label{fig1}}
\end{figure}


\begin{figure}
\epsscale{.65}
\plotone{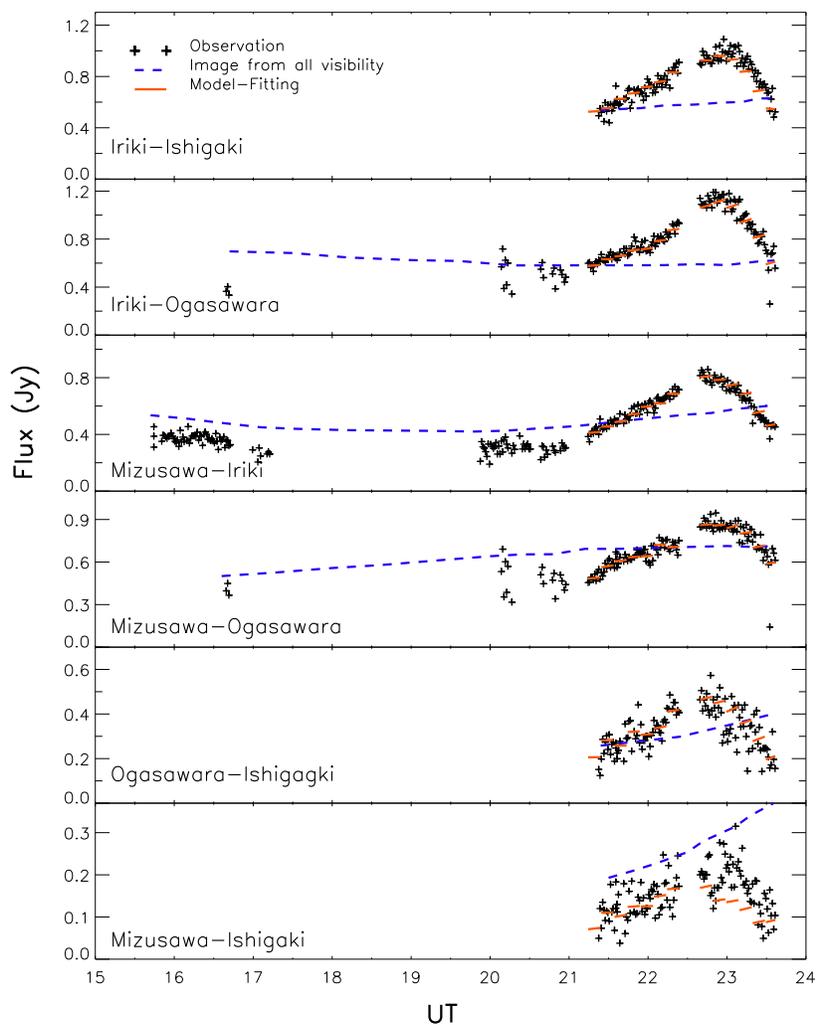}
\caption{Visibility amplitude for the 2007 May 29 observation.
The six baselines are plotted in the order of increasing baseline length
from the top to bottom panels.
The observed visibility flux densities for all baselines are shown,
shown by the black crosses.
During 21$-$24 UT, a mini-flare was detected in all the baselines.
   The blue dashed lines are for the visibility amplitude
based on the image produced from the visibilities
from all the observing times.
   The orange solid lines for each time bin of 10 minutes are
model fits based on a single elliptical Gaussian component.
The baseline length ranges from 1091 to 2270 km$^1$.
}
\label{Fig 2}
\end{figure}
\clearpage

\begin{figure}
\epsscale{0.75}
\plotone{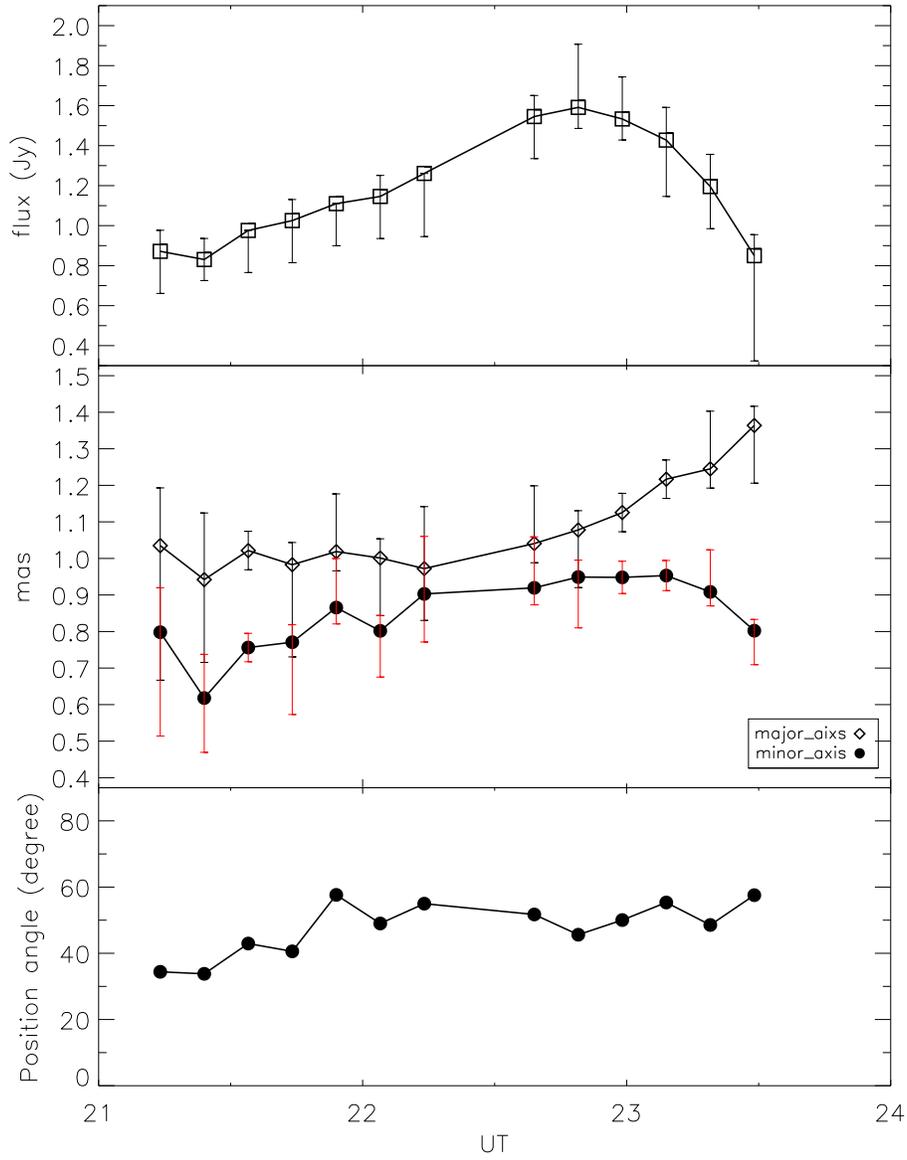}
\caption{Total flux density (top),
sizes of major and minor axes (middle) and position angle (bottom panel)
of model fits based on a single elliptical Gaussian component
with a time bin of 10 minutes during the mini-flare.
   The error bars for the position angle are smaller than the black filled circles.
(A color version of this figure is available in the online journal.)
\label{fig3}}
\end{figure}

\end{document}